\documentclass[pra,aps,twocolumn,showpacs,10pt]{revtex4-1}
\usepackage{amsmath,amssymb,graphicx,amsthm,dsfont,bm,color,ulem}

\begin{document}

\title{General laws of the propagation of ultrafast vortices in free space}

\author{Miguel A. Porras and Ra\'ul Garc\'{\i}a-\'Alvarez}

\affiliation{Grupo de Sistemas Complejos, ETSIME, Universidad Polit\'ecnica de Madrid, Rios Rosas 21, 28003 Madrid, Spain}

\begin{abstract}
We conduct a theoretical study of the propagation of few-cycle, ultrafast vortices (UFVs) carrying orbital angular momentum (OAM) in free space. Our analysis reveals much more complex temporal dynamics than that of few-cycle fundamental Gaussian-like beams, particularly when approaching the single-cycle regime and the magnitude of the topological charge $l$ is high. The recently described lower bound $\sqrt{|l|}$ to the number of oscillations of UFVs with propagation-invariant temporal shape (isodiffracting UFVs) is found to hold on average also for UFVs of general type, with variations along the propagation direction above and below that bound, even vanishing locally. These variations are determined by the so-called Porras factor or $g_0$-factor characterizing the dependence of the Rayleigh distance of the spectral constituents with frequency. With a given available bandwidth, UFVs must widen temporally with increasing magnitude of the topological charge, and must widen or may shrink temporally during propagation as a result of the axially varying, $g_0$-dependent lower bound. Under very restrictive conditions in their generation, an UFV can be shrunk below the lower bound $\sqrt{|l|}$ at a focus into a kind of locally compressed state of OAM, but it broadens well-above $\sqrt{|l|}$ and distorts in a tiny fraction of the depth of focus because of the dispersions introduced by Gouy's phase and wave front mismatch. These propagation phenomena have implications and should be taken into account in experiments and applications of UFVs, such as the generation of high-harmonics and attosecond pulses with high OAM, or in OAM-based ultrafast communications systems, as well as in other areas of physics such as acoustics or electron waves.
\end{abstract}

\maketitle

\section{Introduction}

\noindent In recent years there has been a great deal of interest in the generation and applications of ultrafast vortices (UFVs), or ultrashort pulses carrying orbital angular momentum (OAM), particularly those of with few optical cycles approaching the single-cycle regime, and in their application to high-harmonic and attosecond pulse generation, high-resolution imaging, and fast optical classical and quantum information processing, among others. In an important part of the experiments the aim is to generate shorter and shorter vortices of better quality by eliminating undesirable effects such as topological charge and angular dispersions \cite{BEZUHANOV,MARIYENKO,BEZUHANOV2,ZEYLIKOVICH,TOKIZANE,SHVEDOV,RICHTER,YAMANE,BOCK,APURV,MIRANDA,GRUNWALD,NAIK}. Technical improvements have made it possible to approach the single-cycle regime of durations and to reach two-digit topological charges of the vortices \cite{APURV}. On the other hand, the use of these UFVs in strong-field light-matter interactions, as in high-harmonic and attosecond generation experiments \cite{HERNANDEZ,GARIEPY,REGO,GENEAUX,GAUTHIER,KONG,HERNANDEZ2,DORNEY}, has led to the generation of UFVs of high topological charges, typically a few dozen, even exotic waves with fractional and time varying topological charges \cite{TURPIN,REGO2}. Recent proposals \cite{ZHANG} that mimic  high-harmonic generation based on phase-only spatial light modulators allow for increased topological charge on demand.

Given these huge experimental efforts, it is somewhat surprising the little amount of theoretical work on the propagation characteritics of UFVs \cite{FENG,PARIENTE,LECKNER,NIE,NIE2,PORRAS1,PORRAS2,PORRAS3,PORRAS4,PORRAS5,CONTI1,CONTI2}, even in the a priori simplest situation of free-space propagation. A comprehensive theory of the propagation dynamics of UFVs, similar to that developed a few decades ago for fundamental, few-cycle pulses without OAM \cite{AKTURK,SHEPPARD,KAPLAN,PORRAS6,PORRAS7,PORRAS8,PORRAS9,PORRAS10} when the single-cycle regime was reached \cite{BRABEC}, is still pending.

Indeed the propagation dynamics of UFVs differs in substantial aspects from that of fundamental pulses without OAM because of a strong coupling between the temporal and OAM degrees of freedom in the UFV, as reported in \cite{PORRAS1,PORRAS2,PORRAS3,PORRAS4,PORRAS5} for Laguerre-Gauss-type UFVs and in \cite{PORRAS5,CONTI1,CONTI2} for nondiffracting X-type vortices. The effects of this coupling remain small, but are observable, at low topological charges and/or many-cycle durations, but become large and dominate the propagation dynamics when the single-cycle and high topological charge regimes are approached.

These previous studies considered the so-called isodiffracting UFVs, characterized by a Rayleigh distance that is independent of the frequency of the monochromatic Laguerre-Gauss (LG) constituents. Isodiffracting UFVs play a central role in the theory of UFV propagation because they are the only UFVs whose pulse temporal shape do not change during propagation regardless how short the pulse and how high the topological charge are, as pointed out in \cite{PORRAS1}. For these UFVs, \cite{PORRAS1} establishes a strong coupling between the pulse temporal shape at the bright ring surrounding the vortex singularity and the magnitude of the topological charge that settles an upper bound to the topological charge that an UFVs of certain number of oscillations can carry, and vice versa, settles a lower bound, $\sqrt{|l|}$, to the number of oscillations of an UFV of given magnitude of the topological charge, $|l|$. As a result, an UFV synthesized with a certain bandwidth must increase its duration from that expected from its bandwidth when the imprinted topological charge is increased \cite{PORRAS2}.

However, femtosecond laser sources that emit with frequency independent Rayleigh distance seem to be more the exception than the rule \cite{HOFF,ZHANG2}. The dependence of the Rayleigh distance on frequency is characterized by the $g_0$-factor, first introduced in \cite{PORRAS10}. For ultrafast Gaussian beams, a small red or blue carrier frequency shift experienced during propagation is determined by its $g_0$-factor. More importantly, the $g_0$-factor also determines the carrier-envelope phase distribution in the focal volume \cite{PORRAS10,PORRAS11}. This is why the $g_0$-factor of the source has proved to be a crucial parameter in phase-sensitive light-matter interactions, as in \cite{HOFF,ZHANG2} for electron photoemission, and in \cite{JOLLY} for electron acceleration with radially polarized pulses, since the outcome of experiments depend significantly on the $g_0$-factor. Vice versa, use of phase-sensitive interactions allow the measurement of the carrier-envelope phase map, and from it determine, the $g_0$ factor of the source \cite{HOFF}. The value $g_0=0$ corresponds to the isodiffracting geometry in ultrafast Gaussian beams and UFVs, but the above measurements yielded values of $g_0$ between $-1$ and $-2$, implying an important variation of the Rayleigh distance with frequency.

In this paper we study the propagation (focusing) of UFVs of the LG type beyond the isodiffracting model, and therefore those synthesized from actual femtosecond laser sources with $g_0$-factors different from zero. We do not intend to describe in detail the propagation features of the different types of UFVs, whose detailed description could be deferred to separate studies, but reveal general laws in the form of temporal-OAM couplings that affect all them and underlie the different propagation phenomena observed numerically. For clarity, we often compare the new phenomena with those already known in ultrafast Gaussian beams and isodiffracting UFVs. We find that the lower bound to the number of oscillations of the pulse at the bright ring proportional to $\sqrt{|l|}$ continues to hold for general UFVs, with upward and downward axial variations averaging in $\sqrt{|l|}$ or a higher value whose location is dictated by the $g_0$-factor of the source. Thus, as for isodiffracting UFVs, the duration of general UFVs increases compared to that expected from the available source bandwidth with increasing imprinted topological charge. Unlike ultrafast Gaussian beams and isodiffracting UFVs, the duration and shape of general UFVs changes during propagation as a result of the axially varying lower bound, these variations being more pronounced as $|l|$ and $|g_0|$ are larger, and approaching the single-cycle regime. At certain axial locations it is possible to diminish the number of oscillations below $\sqrt{|l|}$, but this is only feasible in practice with sources with $0<g_0\le 1$ at the far field and sources with $-1\le g_0<0$ at the waist or focus, the optimum situation is the use of a source with $g_0=-1$. Even if it is possible to locally beat the $\sqrt{|l|}$ limit in a kind of compressed state of OAM, the UFV broadens to a number of oscillations well-above $\sqrt{|l|}$ in a small fraction of the depth of focus because the strong dispersive effects of Gouy's phase and wave front mismatch in UFVs with $g_0\neq 0$ and high topological charges.

\section{Preliminaries on ultrafast Laguerre-Gauss vortices}

We represent an UFV of topological charge $l$ propagating in free space as the superposition
\begin{equation}\label{F}
E(r,t',z)e^{il\varphi}=\frac{1}{\pi} \int_0^{\infty}\hat E(r,\omega,z) e^{-i\omega t'} d\omega e^{il\varphi}
\end{equation}
of LG monochromatic light beams
\begin{eqnarray}\label{E}
\hat E(r,\omega,z)&=& \hat a(\omega) \frac{s(\omega)}{s(\omega,z)} \left[\frac{\sqrt{2}r}{s(\omega,z)}\right]^{|l|} e^{-\frac{r^2}{s^2(\omega,z)}} \nonumber \\
&\times &  e^{i\frac{\omega r^2}{2cR(\omega,z)}} e^{-i(|l|+1)\psi(\omega,z)}
\end{eqnarray}
of zero radial order (the only ones used in the cited experiments) and of the same charge $l$. In the above equations, $z$ is the paraxial propagation direction, $(r,z,\varphi)$ are cylindrical coordinates, and $t'=t-z/c$ is the local time. Also, $s(\omega)=\sqrt{2z_R(\omega)c/\omega}$ is the waist Gaussian width,
\begin{equation}\label{SZ}
s(\omega,z)=s(\omega)\sqrt{1+\frac{z^2}{z^2_R(\omega)}}=\sqrt{\frac{2z_R(\omega)c}{\omega}}\sqrt{1+\frac{z^2}{z^2_R(\omega)}}
\end{equation}
is the Gaussian width at each distance, $R(\omega,z)= z + z_R^2(\omega)/z$ is the radius of curvature of the wave fronts, $\psi(\omega,z)= \tan^{-1}[z/z_R(\omega)]$ is Gouy's phase of the fundamental Gaussian beam, and $z_R(\omega)$ is the Rayleigh distance. The divergence angle at the far field can be evaluated from $\theta(\omega) = \sqrt{2c/\omega z_R(\omega)}$. Being limited to positive frequencies, the optical field $E$ in (\ref{F}) is the analytical signal complex representation of the real optical field $\mbox{Re}\{E\}$ \cite{BORNWOLF}.

We also consider the spatial distribution of pulse energy, energy density, or fluence, given by
\begin{eqnarray}\label{FL}
{\cal E}(r,z) &=&\int_{-\infty}^\infty \!\![\mbox{Re} E(r,t',z)]^2 dt' = \frac{1}{2}\int_{-\infty}^\infty \!\!|E(r,t',z)|^2 dt' \nonumber \\
&=& \frac{1}{\pi}\int_0^\infty \!\!|\hat E(r,\omega,z)|^2 d\omega\,,
\end{eqnarray}
which vanishes at the vortex center at $r=0$ and at infinity for a localized field, and then takes a maximum value at a certain radius
$r_{\rm max}$ at each propagation distance, referred henceforth to as the radius $r_{\rm max}$ of the bright ring. The temporal shape of the pulse at this radius is particularly relevant in experiments, specially those involving nonlinear propagation and interactions with matter.

The optical field in Eqs. (\ref{F}) and (\ref{E}) may be regarded as generated on a plane source at $z=0$. Alternatively, and more closely related to current experiments, Eqs. (\ref{F}) and (\ref{E}) also represent the focused optical field in the Debye approximation (no focal shift) \cite{PORRAS10}, with focus at $z=0$, when an ideal focusing element of focal length $f$, as a spherical or parabolic mirror, is illuminated by an input field in the form of a collimated UFV
\begin{equation}\label{LENS1}
E_L(r,t)e^{il\varphi} = \frac{1}{\pi}\int_0^{\infty} \hat E_L(r,\omega)e^{-i\omega t} d\omega e^{il\varphi}
\end{equation}
made of the collimated monochromatic LG beams
\begin{equation}\label{LENS2}
\hat E_L(r,\omega)e^{il\varphi} = \hat A(\omega) \left[\frac{\sqrt{2}r}{S(\omega)}\right]^{|l|}e^{-r^2/S^2(\omega)}e^{il\varphi}
\end{equation}
of Rayleigh distance $Z_R(\omega)$, Gaussian width $S(\omega)= \sqrt{2Z_R(\omega)c/\omega}$ on the focusing system, and divergence angle $\Theta(\omega)= \sqrt{2c/\omega Z_R(\omega)}$. We cannot enter on the sophisticated experimental techniques for the generation of vortices of femtosecond duration as in Eq. (\ref{LENS1}) and (\ref{LENS2}), but the Rayleigh distance $Z_R(\omega)$ and the spectrum $\hat A(\omega)$ in these equations are closely related to the geometry and spectrum of the femtosecond laser source. The spectrum $\hat A(\omega)$ is characterized by a certain mean frequency
\begin{equation}
\omega_0= \frac{\int_0^\infty |\hat A(\omega)|^2 \omega d\omega}{\int_0^\infty |\hat A(\omega)|^2 d\omega}\,,
\end{equation}
and corresponds in time domain to a certain pulse shape
\begin{equation}
A(t) = \frac{1}{\pi} \int_0^\infty \hat A(\omega) e^{-i\omega t} d\omega\,
\end{equation}
of a physically meaningful carrier frequency $\omega_0$ if, according to the standard definition \cite{BRABEC}, the full width at half maximum (FWHM) of $|A(t)|^2$ comprises at least one carrier period $2\pi/\omega_0$.

In the Debye approximation of focusing, the Rayleigh distance and spectra of the focused UFV and of the input UFV from the femtosecond source are related by \cite{TACHE}
\begin{equation}\label{DEBYE}
z_R(\omega)= \frac{f^2}{Z_R(\omega)}\,, \quad \hat a(\omega) = -i\frac{f}{z_R(\omega)}\hat A(\omega) \,.
\end{equation}

For ulterior use, given a function $f(\omega)$ of frequency we introduce the notation
\begin{equation}
\overline{f(\omega)} = \frac{\int_0^\infty |\hat E(r,\omega,z)|^2 f(\omega) d\omega}{\int_0^\infty |\hat E(r,\omega,z)|^2 d\omega}
\end{equation}
for its mean value with the spectral density $|\hat E(r,\omega,z)|^2$ of the UFV, and for any function $g(t')$ of time, the notation
\begin{equation}
\overline{g(t')} = \frac{\int_{-\infty}^\infty |E (r,t',z)|^2 g(t') d t'}{\int_{-\infty}^\infty |E(r,t',z)|^2 d t'}
\end{equation}
for its mean value with the intensity $|E(r,t',z)|^2$ of the UFV. Note that these mean values depend in general on $r$ and $z$ because $|\hat E(r,\omega,z)|^2$ and $|E(r,t',z)|^2$ depend on $r$ and $z$ [and because $f(\omega)$ and $g(t')$ may be functions of $r$ and $z$ too]. The variance of $f(\omega)$ is $\sigma^2_{f(\omega)} = \overline{[f(\omega)- \overline{f(\omega)}]^2}=\overline{f^2(\omega)}- \overline{f(\omega)}^2$, and similarly for a function of time. In particular, $\bar\omega$ is the mean or carrier frequency at any point of the UFV, and $\sigma_{\omega}^2 = \overline{\omega^2}- \bar\omega^2$ is the variance of $\omega$. Similarly, $\overline{t'}$ is the mean temporal location, and $\sigma_{t'}= \overline{t^{\prime 2}}- \overline{t'}^2$ is the variance of the pulse intensity. Suitable measures of the spectral bandwidth and duration of the UFV at a given point $(r,z)$ are the so-called Gaussian-equivalent half-bandwidth and half-duration, $\Delta \omega = 2\sigma_\omega$ and $\Delta t=2\sigma_{t'}$, yielding the $1/e^2$-decay half-width for Gaussian spectral density and intensity. The product $\Delta t\Delta \omega$ is always larger or equal to two, with the minimum value of two reached for Gaussian-shaped spectral density and intensity. Also, the product $\bar\omega\Delta t/\pi=2 \Delta t/T$, where $T=2\pi/\bar\omega$ is the mean or carrier period, is the number of oscillations in the full Gaussian-equivalent duration $2\Delta t$, but to avoid these $\pi$ factors in relevant formulas we will refer to $\bar\omega \Delta t$ as the number of oscillations.


\subsection{The $g_0$-factor of the source}

In previous theoretical studies on UFVs, $Z_R(\omega)$, and hence $z_R(\omega)$, are taken to be independent of frequency in the isodiffracting model of UFVs, in which case the temporal shape of the UFV does not change with propagation distance. While this property confers on isodiffracting UFVs a prominent place from the theoretical point of view, current femtosecond laser sources emit pulses with different Rayleigh distances for different frequencies, as recently demonstrated \cite{HOFF,ZHANG2}. Since the function $Z_R(\omega)$ is generally unknown, several simple models are often used \cite{JOLLY}. For example, the model $Z_R(\omega)=Z_R(\omega_0)(\omega/\omega_0)^{g_0}$ yields, using the above relations between $Z_R(\omega)$, $S(\omega)$ and $\Theta(\omega)$, yields $S(\omega) = S(\omega_0)(\omega/\omega_0)^{(g_0-1)/2}$ and $\Theta(\omega)=\Theta(\omega_0)(\omega_0/\omega)^{(g_0+1)/2}$ for the input UFV, and using Eqs. (\ref{DEBYE}), $z_R(\omega)=z_R(\omega_0)(\omega_0/\omega)^{g_0}$, $s(\omega)=s(\omega_0)(\omega_0/\omega)^{(g_0+1)/2}$ and $\theta(\omega)=\theta(\omega_0)(\omega/\omega_0)^{(g_0-1)2}$ for the focused UFV. In particular, $g_0=0$ is the isodiffracting model, $g_0=1$ describes an input UFV with constant width, and hence focused UFV with constant convergence angle and focal width inversely proportional to frequency, and  $g_0=-1$ describes an input UFV with constant divergence angle and width inversely proportional to frequency, corresponding to a focused UFV with convergence angle inversely proportional to frequency and constant width at focus. With other $g_0$-values, none of the parameters are constant. The above formula for $Z_R(\omega)$ is a simple model that will be used in the examples below, but obviously real sources do not have to conform to it.

Fortunately, for pulses with at least one carrier oscillation, it has been theoretically suggested and experimentally demonstrated that it is only the variation of the Rayleigh distance with frequency in the vicinity of the carrier frequency $\omega_0$ that determines most of propagation properties of pulsed beams \cite{PORRAS10,HOFF,PORRAS11}. The different situations are suitably described by a single dimensionless parameter called the $g_0$-factor defined as \cite{PORRAS10}
\begin{equation}\label{G0}
g_0= \left.\frac{dZ_R(\omega)}{d\omega}\right|_{\omega_0} \frac{1}{Z_R(\omega_0)}\omega_0 =
-\left.\frac{dz_R(\omega)}{d\omega}\right|_{\omega_0} \frac{1}{z_R(\omega_0)}\omega_0\,,
\end{equation}
and that characterizes the variation of the Rayleigh range with frequency about the carrier frequency $\omega_0$. The symbol $g_0$ is the same as in the model in the preceding paragraph because they coincide in that model. Thus, $g_0=1$ means constant width $S(\omega)$ only about $\omega_0$, $g_0=-1$ constant divergence about $\omega_0$, and so on. Recent studies outline the need of measuring the $g_0$-factor for each particular femtosecond laser source since its value has been evidenced to strongly affect the outcome of experiments with these sources, particularly those involving phase-sensitive light-matter interactions \cite{HOFF,ZHANG2,JOLLY}. Recent measurements yield values of $g_0$ between $-1$ and $-2$ for femtosecond sources using hollow-core fiber-compressors \cite{HOFF}, but the value of $g_0$ for other source types remains unknown. We assume here that $|g_0|$ does not exceed $2$.

\begin{figure}
\centering
\includegraphics[width=8.5cm]{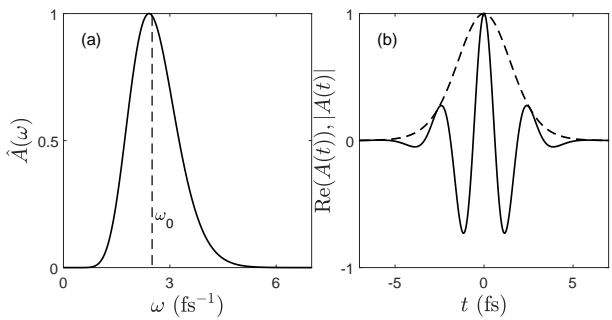}
\caption{\label{Fig1} Power-exponential spectrum $\hat A(\omega)\propto (\omega/\omega_0)^{\alpha-1/2}\exp(-\alpha\omega/\omega_0)$, with $\alpha=14.25$ and $\omega_0=2.5$ rad/fs of the quasi-Gaussian pulse $A(t)=[-i\alpha/(\omega_0 t - i\alpha)^{\alpha+1/2}]$ of carrier frequency $\omega_0$. The value of $\alpha$ is chosen such that the FWHM duration of $|A(t)|^2$ is one carrier period.}
\end{figure}

\subsection{Previous results and open problems}

\begin{figure*}[t!]
\centering
\includegraphics[width=12cm]{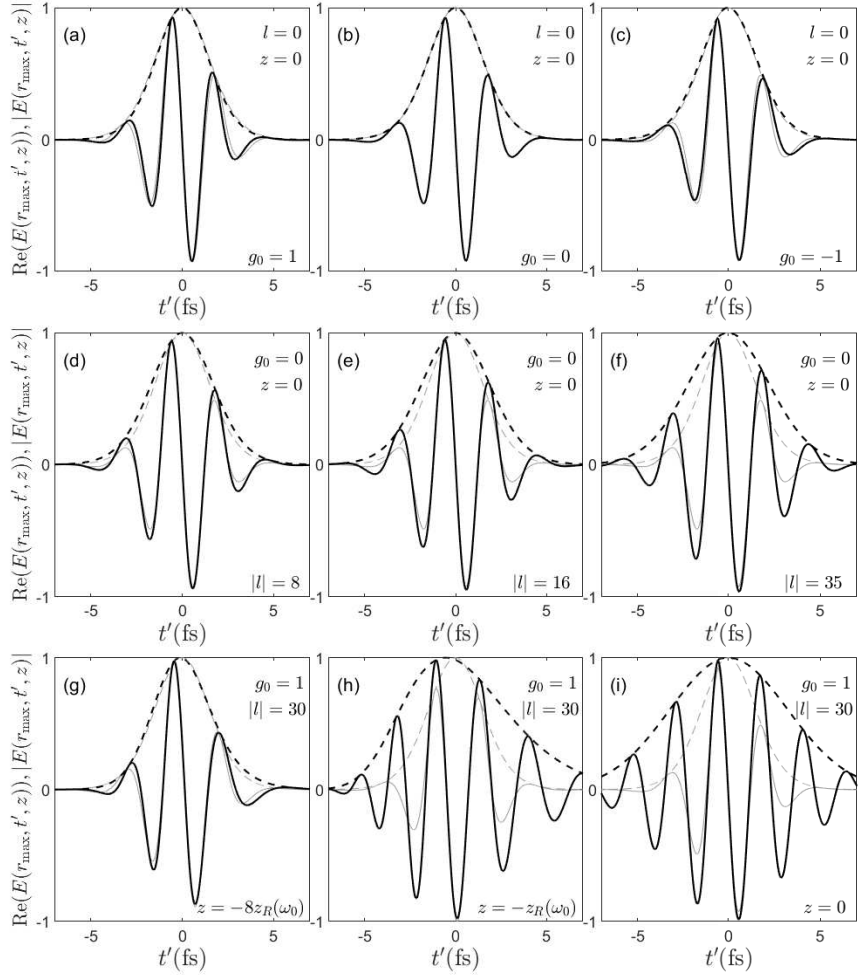}
\caption{\label{Fig2} Focused pulse shapes and amplitudes at the radius $r_{\rm max}$ of maximum fluence for different input optical field of interest. All them are evaluated numerically from Eqs. (\ref{F}), (\ref{E}), (\ref{DEBYE}) with $f=20$ cm, and with the same spectrum $\hat A(\omega)$ of the single-cycle pulse in  Fig. \ref{Fig1}, and using the model $Z_R(\omega)=Z_R(\omega_0)(\omega/\omega_0)^{g_0}$ with $Z_R(\omega_0)=4$ m. (a-c) Pulse shapes at focus for the fundamental pulsed Gaussian beam ($l=0$) with the three choices $g_0=1,0,-1$. All three are undistorted quasi-Gaussian pulses of the same duration during the whole propagation with slightly blue ($g_0=1$) or red shifted frequencies ($g_0=-1$) at focus. (d-f) Pulse shapes at focus in the isodiffracting model with $g_0=0$ and with different values of $|l|$. All three propagate undistorted without any frequency shift but with duration increasing with $|l|$. (g-i) Pulse shapes for $g_0=1$ and $|l|=30$ at different propagation distances. The pulse is slightly blue shifted, distorted and broadened during focusing. The gray curves are always the pulse in Fig. \ref{Fig1}(b) with an artificially added phase in each case for a better appreciation of the frequency shifts. In all figures the peak amplitude is set to unity and shifted to $t'=0$ for a better comparison.}
\end{figure*}

Figures \ref{Fig1} and \ref{Fig2} illustrate previously known results for Gaussian beams and for isodiffracting UFVs, faced to the new phenomena in the propagation of general UFVs considered here. In all cases represented in Fig. \ref{Fig2} the source spectrum $\hat A(\omega)$, shown in Fig. \ref{Fig1}(a), is the same, and corresponds in time domain to the approximately Gaussian pulse $A(t)$ shown in Fig. \ref{Fig1}(b) containing a single oscillation in the FWHM of $|A(t)|^2$. All the graphs in Fig. \ref{Fig2} represent pulse shapes $\mbox{Re}\,E$ and amplitudes $|E|$ of focused UFVs at the radii $r_{\rm max}$ of maximum fluence at each propagation distance.

First, focusing of few-cycle, even single-cycle pulses in the form of a fundamental Gaussian beam [$l=0$ in Eqs. (\ref{F}) and (\ref{E}) and $r_{\rm max}=0$] only introduces minor changes on the pulse temporal shape [Figs. \ref{Fig2}(a), (b) and (c)], irrespective of the particular value of $g_0$, i. e., the pulse continues to be an approximately single-cycle, Gaussian-like pulse during the whole propagation \cite{PORRAS8}, with slightly blue shifted frequency for $g_0>0$, and red shifted frequency for $g_0<0$ about the focus \cite{SHEPPARD,KAPLAN,PORRAS10}, and with different maps of carrier-envelope phase in the focal volume, as studied in detail in \cite{PORRAS10,PORRAS11}.

Second, with the same source spectrum in Fig. \ref{Fig1} of a single-cycle pulse, the number of oscillations of the synthesized isodiffracting input UFV ($g_0=0$) monotonically increases with the magnitude of the topological charge from those expected from the source pulse $A(t)$, no matter by what technical means it is generated, and the focused isodiffracting UFV maintains this temporal shape during the whole focusing process, with no appreciable blue or red frequency shift at the bright ring, $\bar\omega\simeq \omega_0$. The increase of the number of oscillations with $|l|$ is a consequence of the upper bound to the relative bandwidth $\sigma_\omega/\bar\omega< 1/\sqrt{|l|}$ at the bright ring described in \cite{PORRAS1}, or on account of relation $\Delta t\Delta\omega\ge 2$ and $\Delta\omega=2\sigma$, the lower bound $\bar\omega \Delta t>\sqrt{|l|}$ to the number of oscillations \cite{PORRAS1}. In addition, with $\bar\omega\simeq \omega_0$ for isodiffracting UFV \cite{PORRAS5}, $\Delta t>\sqrt{|l|}/\omega_0$ imposes directly a lower bound to the pulse duration. Thus, irrespective of how wide is the source spectrum, or how short is the source pulse $A(t)$ that can be synthesized with it, the UFV adapts itself to a number of oscillations satisfying $\omega_0\Delta t >\sqrt{|l|}$ \cite{PORRAS2}.

The intention of this paper is to understand propagation phenomena of UFVs with $g_0\neq0$, such as those observed in Figs. \ref{Fig2}(g), (h) and (i). In this example $g_0=1$ because the width $S(\omega)\equiv S(\omega_0)$ of the input UFV is taken to be independent of frequency, and according to Eqs. (\ref{LENS1}) and (\ref{LENS2}) the input field is $E_L(r,t)e^{il\varphi}= A(t)[\sqrt{2}r/S(\omega_0)]^{|l|} e^{-r^2/S^2(\omega_0)} e^{il\varphi}$. First, and unlike pulsed Gaussian beams and isodiffracting UFVs, the pulse shape at $r_{\rm max}$ changes during focusing. Pulse distortion is weak for small $|l|$ but quite pronounced for large $|l|$, as in Figs. \ref{Fig2}(g), (h) and (i). Second, the topological charge and duration of the input UFV in Fig. \ref{Fig1}(b), and in its initial stage of focusing in Fig. \ref{Fig2}(g), are chosen so that inequality $\omega_0\Delta t>\sqrt{|l|}$ is violated, contradicting apparently the results in \cite{PORRAS1}. However, this lower bound applies only to the most fundamental situation of UFVs with propagation-invariant pulse shape. Indeed, there is no restriction, on physical grounds, to produce the space-time factorized field $E_L(r,t)e^{il\varphi}= A(t)[\sqrt{2}r/S(\omega_0)]^{|l|}e^{-r^2/S^2(\omega_0)}e^{il\varphi}$ with $A(t)$ as short and $|l|$ as large as desired at a given transversal plane, only technical issues. About the focus, however, the UFV is distorted and broadened, as in Figs. \ref{Fig2}(h) and (i), so that inequality $\bar\omega\Delta t>\sqrt{|l|}$ is satisfied by far. In the following sections we demonstrate that this behavior is a result of more general restrictions to the number of oscillations at the bring ring that generalize $\bar\omega \Delta t>\sqrt{|l|}$ for isodiffracting UFVs to general UFVs.

\section{General restrictions to the pulse properties at the bright ring}

To that purpose we first locate the maximum of the fluence distribution at each transversal plane. Differentiating with respect to $r$ the fluence in Eq. (\ref{FL}) with the spectral density $|\hat E(r,\omega,z)|^2$ obtained from Eq. (\ref{E}), we obtain, after some algebra,
\begin{equation}
\frac{d{\cal E}}{dr}= \frac{2}{r}\int_0^\infty d\omega |\hat E|^2 \left[|l|- \frac{2r^2}{s_\omega^2(z)}\right] \, ,
\end{equation}
which equated to zero leads to the implicit equation
\begin{equation}
\frac{|l|}{2} = r_{\rm max}^2 \overline{\left( \frac{1}{s^2(\omega,z)}\right)}(r_{\rm max})
\end{equation}
for the maxima or minima of the fluence, where we have explicitly written that the $(r,z)$-dependent mean value $\overline{1/s^2(\omega,z)}$ is evaluated at the radius of $r_{\rm max}$ of the bright ring. Differentiating again with respect to $r$, the second derivative yields a cumbersome and long expression, which evaluated again at $r_{\rm max}$ yields however the simpler expression
\begin{eqnarray}\label{SECOND}
\left.\frac{d^2{\cal E}}{dr^2}\right|_{r_{\rm max}} &=& -8{\cal E}(r_{\rm max})\overline{\left( \frac{1}{s^2(\omega,z)}\right)}(r_{\rm max}) \nonumber \\
&\times& \left[|l|+1-|l| \frac{\overline{1/s^4(\omega,z)} (r_{\rm max})}{\overline{1/s^2(\omega,z)}^2(r_{\rm max})}\right]
\end{eqnarray}
where as above the mean values are explicitly written to be evaluated at $r_{\rm max}$. The condition of maximum of the fluence in Eq. (\ref{SECOND}) then leads to the inequality
\begin{equation}
\frac{\overline{1/s^4(\omega,z)}(r_{\rm max})}{\overline{1/s^2(\omega,z)}^2(r_{\rm max})} < \frac{|l|+1}{|l|}\,,
\end{equation}
or equivalently, to the inequality
\begin{equation}\label{UPPER}
  \frac{\sigma^2_{1/s^2(\omega,z)}}{\overline{1/s^2(\omega,z)}^2} = \frac{\overline{1/s^4(\omega,z)}- \overline{1/s^2(\omega,z)}^2}{\overline{1/s^2(\omega,z)}^2} < \frac{1}{|l|}\,,
\end{equation}
that is satisfied by any UFV at its maximum of fluence at any propagation distance. In equality (\ref{UPPER}) we have omitted again
$r_{\rm max}$ in the mean values and variance to lighten the notation, but it should be understood from now on that they are evaluated at this radius. Inequality (\ref{UPPER}) is the main mathematical result of this paper, stating that the relative variance of the function of frequency $1/s^2(\omega,z)$ at the radius of maximum fluence of general UFVs is restricted by the upper bound in the right hand side (r.h.s.) of inequality (\ref{UPPER}). The physical interpretation and consequences of this restriction is the purpose of the remainder of this paper.

Although inequality (\ref{UPPER}) holds for arbitrarily short UFVs, i. e., also for sub-cycle pulses of arbitrary temporal shape and ultrabroadband spectrum, we limit from now our considerations to pulses with at least one carrier oscillation as defined in \cite{BRABEC}, with relatively narrow spectrum, as in Fig. \ref{Fig1}, and therefore with a physically meaningful carrier frequency. With this limitation inequality (\ref{UPPER}) can be transformed under suitable approximations into a practical inequalities involving the carrier frequency, the bandwidth, and duration of the UFVs.

\section{Red and blue frequency shifts}

To this end, we first investigate about the actual carrier frequency of the oscillations of UFVs at their bright ring. It has been demonstrated in \cite{PORRAS5} that the carrier frequency of isodiffracting UFVs is no appreciably shifted from the source frequency $\omega_0$, that is, $\bar\omega\simeq \omega_0$, in line with what happens to the fundamental isodiffracting pulsed Gaussian beam \cite{PORRAS6}. For UFVs with $g_0\neq 0$ there are significant, but not large, blue or red shifts of the carrier frequency, which are also similar to those of fundamental pulsed Gaussian beams of the same value of $g_0$ \cite{PORRAS10}, and are substantially independent of the topological charge.

\begin{figure}
\centering \includegraphics[width=9cm]{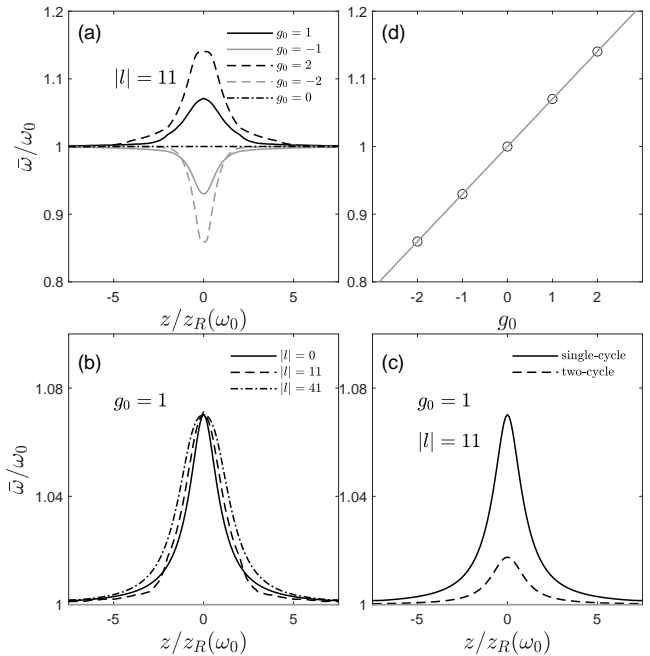} \caption{\label{Fig3} Frequency shifts of UFVs at their bright ring. They are evaluated  numerically with Eqs. (\ref{F}), (\ref{E}), (\ref{DEBYE}), with $Z_R(\omega)=Z_R(\omega_0)(\omega/\omega_0)^{g_0}$ and $\hat A(\omega)= (\omega/\omega_0)^{\alpha-1/2}\exp(-\alpha\omega/\omega_0)$ with $\alpha=14.25$ (single-cycle pulse), except in (d) where $\alpha=14.25$ for the single-cycle pulse and $\alpha=57.11$ for the two-cycle pulse. Frequency shift (a) for several values for $g_0$, (b) for different values of $|l|$, and (c) for different pulse durations, all them as functions of propagation distance. (d) Frequency shifts at focus $z=0$ in (a) as a function of $g_0$, fitting approximately a stright line.}
\end{figure}

In Fig. \ref{Fig3}(a) the carrier frequency $\bar\omega$ at the bright ring is represented for UFVs of the same topological charge and different values of $g_0$, versus propagation distance $z$ about the focus for a single-cycle, approximately Gaussian source pulse $A(t)$ (see caption for details). Starting in all cases with a carrier frequency $\bar\omega\simeq \omega_0$ far from the focus, the carrier frequency is increasingly red shifted approaching the focus for negative $g_0$, and a increasingly blue shifted approaching the focus for positive $g_0$. As seen in Fig. \ref{Fig3}(b) for $g_0=1$, the frequency shift at focus does not appreciably depend on $l$ and is approximately equal to that affecting the fundamental Gaussian beam of the same value of $g_0$. Off-focus the dependence of the frequency shift with the topological charge is also weak. For longer pulses, as for the two-cycle source pulse $A(t)$ in Fig. \ref{Fig3}(c), the frequency shifts are much less pronounced, and vanish in the monochromatic limit, as expected. Thus, the largest red and blue shifts correspond to single-cycle input pulses at focus, and they fit the approximate linear variation with the $g_0$-factor shown in Fig. \ref{Fig3}(d). These maximum frequency shifts are of course relevant in experiments, but for $|g_0|\le 2$ do not exceed a relative variation of 15 \% with respect to the source carrier frequency, which justifies the approximations that will be made in the following section. Although particular power-exponential source spectra, corresponding to approximately Gaussian-shaped input pulses, are used in Fig. \ref{Fig3} (see caption for  details), we have observed in additional numerical simulations with other source spectra similar relative frequency shifts between $10$\% and $20$\%.

\section{Restrictions to the bandwidth and duration}

For $z_R(\omega)$ independent of frequency it can readily be seen from Eq. (\ref{SZ}) that inequality (\ref{UPPER}) reduces to the previously known inequality $\sigma_\omega/\bar\omega < 1/\sqrt{|l|}$ for isodiffracting UFVs that involves physically meaningful properties of the pulse at $r_{\rm max}$. Taking into account that $\Delta\omega=2\sigma_\omega$ and that $\Delta\omega\Delta t \ge 2$, the number of oscillations of isodiffracting UFVs satisfies $\bar\omega\Delta t \ge \sqrt{|l|}$ \cite{PORRAS1}.

Similar, though approximate, restrictions involving the carrier frequency, bandwidth and duration at $r_{\rm max}$ of general UFVs with at least one cycle can be obtained as follows. Using the approximate equality $\sigma^2_{f(\omega)}\simeq [df(\omega)/d\omega|_{\bar\omega}]^2\sigma_\omega^2$ for $f(\omega)=1/s^2(\omega,z)$ frequently used in statistics \cite{HOEF}, evaluating the derivative as $df(\omega)/d\omega = -[1/s^4(\omega,z)]d s^2(\omega,z)/d\omega$ for convenience, and approaching $\overline{f(\omega)}$ in the denominator of (\ref{UPPER}) by the first order Taylor series $\overline{f(\omega)} \simeq \overline{f(\bar\omega)+ df(\omega)/d\omega|_{\bar\omega}(\omega-\bar\omega)}=f(\bar\omega)$, we obtain
\begin{equation}\label{UPPER1}
\sigma_\omega < \frac{1}{\sqrt{|l|}}\left|\frac{s^2(\bar\omega,z)}{\left.\frac{d s^2(\omega,z)}{d\omega}\right|_{\bar\omega}}\right|\,.
\end{equation}
Introducing the bandwidth $\Delta\omega = 2\sigma_\omega$ and using that $\Delta\omega\Delta t\ge 2$, inequality (\ref{UPPER1}) yields
\begin{equation}\label{LOWER1}
\Delta t > \sqrt{|l|} \left|\frac{\left.\frac{d s^2(\omega,z)}{d\omega}\right|_{\bar\omega}}{s^2(\bar\omega,z)}\right| \,.
\end{equation}
Explicit evaluation of the derivative yields the result
\begin{equation}\label{UPPER2}
\frac{\sigma_\omega}{\bar\omega} < \frac{1}{\sqrt{|l|}}\frac{1}{\left|1+ g(\bar\omega)\displaystyle\frac{1-z^2/z^2_R(\bar\omega)}{1+z^2/z^2_R(\bar\omega)}\right|}\, ,
\end{equation}
and
\begin{equation}\label{LOWER2}
\bar\omega \Delta t > \sqrt{|l|}\left|1+ g(\bar\omega)\displaystyle\frac{1-z^2/z^2_R(\bar\omega)}{1+z^2/z^2_R(\bar\omega)}\right| \, ,
\end{equation}
where $g(\omega)= -[(dz_R(\omega)/d\omega)/z_R(\omega)]\omega$. The r.h.s. of inequality (\ref{LOWER2}) imposes a lower bound to the number of oscillations of the pulse that is different at each propagation distance. Its evaluation is however difficult because one needs to know the functions of frequency $z_R(\omega)$ and $g(\omega)$, and then evaluate them at the carrier frequency $\bar\omega$ at the bright ring at each propagation distance. Experimentally this would require a careful characterization of the input source by determining $Z_R(\omega)$ as a function of frequency, and measuring $\bar\omega$ at the bright ring at each selected distance. In a numerical simulation of an experiment, one would need to specify models of $\hat A(\omega)$ and $Z_R(\omega)$ of the input pulse, use Eqs. (\ref{DEBYE}), compute the focused optical field with Eqs. (\ref{F}) and (\ref{E}), and extracting the values of $\bar\omega$. As seen in the previous section, the carrier frequency $\bar\omega$ at $r_{\rm max}$ may be red or blue shifted with respect to the carrier frequency $\omega_0$ of the source, but this shift does not exceed a relative value of $10$-$20$\% for the extreme case of single-cycle pulses, for $|g_0|\le 2$, with any topological charge, and vanishes as the number of oscillations increase, regardless the particular choice of $\hat A(\omega)$ and $Z_R(\omega)$. Thus, we can transform the upper bound in (\ref{UPPER}) and the lower bound in (\ref{LOWER2}) into approximate, but much easier to evaluate upper and lower bounds by replacing $\bar\omega$ with the source frequency $\omega_0$ in the r.h.s. of inequalities (\ref{UPPER2}) and (\ref{LOWER2}), to obtain
\begin{equation}\label{UPPER3}
\frac{\sigma_\omega}{\bar\omega} < \frac{1}{\sqrt{|l|}}\frac{1}{\left|1+ g_0\displaystyle\frac{1-z^2/z^2_R(\omega_0)}{1+z^2/z^2_R(\omega_0)}\right|}\, ,
\end{equation}
and
\begin{equation}\label{LOWER3}
\bar\omega \Delta t > \sqrt{|l|}\left|1+ g_0\displaystyle\frac{1-z^2/z^2_R(\omega_0)}{1+z^2/z^2_R(\omega_0)}\right| \,,
\end{equation}
whose r.h.s. are determined by standard properties of the source such as its carrier frequency $\omega_0$ and Rayleigh distance $z_R(\omega_0)=f^2/Z_R(\omega_0)$ at the carrier frequency. The appearance of the $g_0$-factor underlines the need to measure it for the available laser source. Inequality (\ref{LOWER3}), supplemented by inequality (\ref{UPPER3}), is the main practical result of this paper that imposes a $z$-dependent lower bound proportional to $\sqrt{|l|}$ to the number of oscillations at the bright ring of general UFVs, and generalizes the $z$-independent lower bound $\sqrt{|l|}$ in the isodiffracting case.

It is possible to derive more intuitively the above results by examining more closely the spectral density
\begin{equation}
|\hat E(r,\omega,z)|^2 = \frac{f^2}{z_R^2(\omega)} \frac{s^2(\omega)}{s^2(\omega,z)}|\hat A(\omega)|^2 \! \left[\frac{2r^2}{s^2(\omega,z)}\right]^{|l|} \!\! e^{-\frac{2r^2}{s^2(\omega,z)}}\,.
\end{equation}
Using the approximate equality  $x^{2m}e^{-x^2}\simeq e^{-2(x-\sqrt{m})^2} (m/e)^m$, which is more accurate as $m>0$ is larger, the spectral density can be approximated by
\begin{equation}
|\hat E(r,\omega,z)|^2 \simeq \frac{f^2}{z_R^2(\omega)} \frac{s^2(\omega)}{s^2(\omega,z)} |\hat A(\omega)|^2 e^{-2\left(\frac{\sqrt{2}r}{s(\omega,z)}-\sqrt{|l|}\right)^2}\,,
\end{equation}
where we have omitted the irrelevant $(m/e)^m$ factor. The two first factors are present also with $l=0$, induce carrier frequency shifts but do not significantly alter the bandwidth of the source spectrum $\hat A(\omega)$, in the same way as for pulsed Gaussian beams. The last factor is peculiar to UFVs and acts as a band-pass filter when $s(\omega,z)$ depends on frequency that limits the bandwidth, and hence the duration. Approximating $1/s(\omega,z)\simeq 1/s(\omega_0,z) + d[1/s(\omega,z)]/d\omega|_{\bar\omega}(\omega-\bar\omega)$, expressing for convenience the derivative as $d[1/s(\omega,z)]/d\omega = -[d s^2(\omega,z)/d\omega]/2s^3(\omega,z)$, and evaluating the spectral density at $r_{\rm max}^2 = (|l|/2)[1/\overline{1/s^2(\omega,z)}]\simeq (|l|/2) s^2(\bar\omega,z)$, one arrives at
\begin{equation}\label{SD2}
|\hat E(r,\omega,z)|^2 \simeq \frac{f^2}{z_R^2(\omega)} \frac{s^2(\omega)}{s^2(\omega,z)} |\hat A(\omega)|^2 e^{-\frac{(\omega-\bar\omega)^2}{2\sigma^2_G}}
\end{equation}
with
\begin{equation}
\sigma_G = \frac{1}{\sqrt{|l|}}\left|\frac{s^2(\bar\omega,z)}{\left.\frac{d s^2(\omega,z)}{d\omega}\right|_{\bar\omega}}\right|\, .
\end{equation}
As the product of $|\hat A(\omega)|^2$ and the last Gaussian factor in Eq. (\ref{SD2}), the spectral density of the UFV at $r_{\rm max}$ cannot by wider than $\sigma_G$, that is, $\sigma_\omega < \sigma_G$, which is the same as inequality (\ref{UPPER1}), from which the remainder of inequalities were derived.

\begin{figure}
\centering \includegraphics*[width=8.5cm]{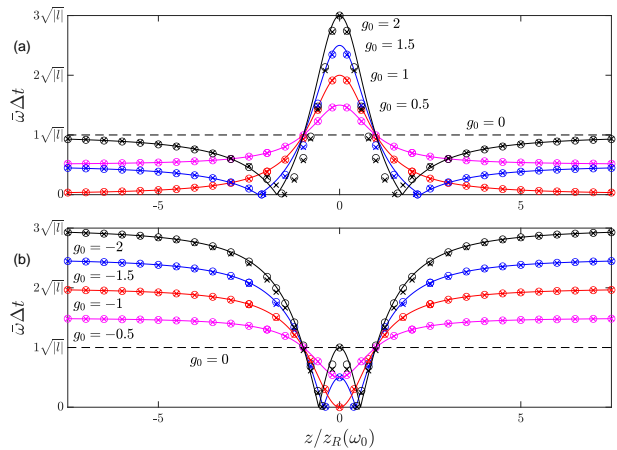}\caption{\label{Fig4} Lower bound to the number of oscillations of UFVs as a function of propagation distance for several (a) positive and (b) negative values of the $g_0$-factor. Solid curves: Lower bound given by the r.h.s. of inequality (\ref{LOWER3}). Symbols: Lower bound evaluated from the r.h.s. of inequality (\ref{LOWER2}) using numerical simulations of the propagation with the source model $Z_R(\omega)=Z_R(\omega_0)(\omega/\omega_0)^{g_0}$ and spectra $\hat A(\omega)$ of single-cycle pulses of shapes $A(t)=[-i\alpha/(\omega_0 t - i\alpha)^{\alpha+1/2}]$, $\alpha=14.25$, $\omega_0=2.5$ rad/fs (open circles), and $A(t)=\mbox{sinc}^2{t/T}e^{-i\omega_0 t}$, $T=3.9$ fs, $\omega_0=2.5$ rad/fs (crosses).}
\end{figure}

The $z$-dependent lower bound to the number of oscillations at the bright ring of UFVs as given by the r.h.s. of inequality (\ref{LOWER3}) is represented in Figs. \ref{Fig4}(a) and (b) by means of solid curves for several values of $g_0$.
For positive $g_0$ the lower bound in the focal region $[-z_R(\omega_0), z_R(\omega_0)]$ is above the bound $\sqrt{|l|}$ for isodiffracting UFVs, outside the focal region is below $\sqrt{|l|}$, and at the edges $\pm z_R(\omega_0)$ of the focal region equals $\sqrt{|l|}$. The opposite happens with negative values of $g_0$. For any $|g_0|$ the maximum lower bound is $(1+|g_0|)\sqrt{|l|}$, reached at the focus for positive $g_0$ and far from the focus for negative $g_0$.

Remarkably, for $|g_0|\ge 1$ there exist isolated axial positions
\begin{equation}
z_b=\pm \sqrt{\frac{g_0+1}{g_0-1}} \, z_R(\omega_0) \,,
\end{equation}
located outside the focal region for $g_0\ge 1$ and within the focal region for $g_0\le -1$, where the lower bound vanishes and therefore there is no restriction to the minimum duration of the UFV. At $z_b$ the lower bound disappears because $d s^2(\omega,z)/d\omega|_{\omega_0}=0$, i. e., the Gaussian width of the monochromatic LG constituents is constant about the carrier frequency, which means, according to Eq. (\ref{SD2}) with $\sigma_G=\infty$, that the spectral density is substantially the same as $|\hat A(\omega)|^2$ of the source, except for small changes associated with the frequency shifts. In practice, the absence of lower bound means that the minimum duration of the UFV is only limited by the source spectrum $\hat A(\omega)$ to the duration of $A(t)$. This finding generalizes to different positions $z_b$ for different values of $|g_0|\ge 1$ the introductory example in Figs. \ref{Fig2}(g), (h) and (i) with $z_b=\pm\infty$, i. e., at the focusing system in the Debye approximation, with $g_0=1$.

Figures \ref{Fig4}(a) and (b) also serve to support the validity of (\ref{LOWER3}) to approximate (\ref{LOWER2}). The symbols in these figures represent the more precise lower bound provided by the r.h.s. of inequality (\ref{LOWER2}), which requires specifying particular models of $Z_R(\omega)$ and $\hat A(\omega)$, as detailed in the caption, computing the focused optical field and its actual carrier frequency $\bar\omega$ at the bright ring at each distance. In these figures dots and crosses correspond to source pulses of different shapes (Gaussian-like, and sinc squared) all containing a single oscillation, in which case the frequency shifts are larger and then the discrepancies of (\ref{LOWER3}) from (\ref{LOWER2}) may be more pronounced. The deviations are however small, and would indeed be inappreciable if, for instance, the symbols are evaluated using multiple-cycle input pulses. These simulations support that the r.h.s. of (\ref{LOWER2}) depends weakly on these fine details of the source, and therefore the lower bound to the number of oscillations can be accurately determined by the analytical formula of the r.h.s. of (\ref{LOWER3}), which is exclusively determined by the three parameters $\omega_0$, $z_R(\omega_0)$ and $g_0$ pertaining the source.

\section{Pulse shape changes upon propagation and with the topological charge, and locally compressed states of orbital angular momentum}

\begin{figure}[!]
\centering \includegraphics*[width=8.5cm]{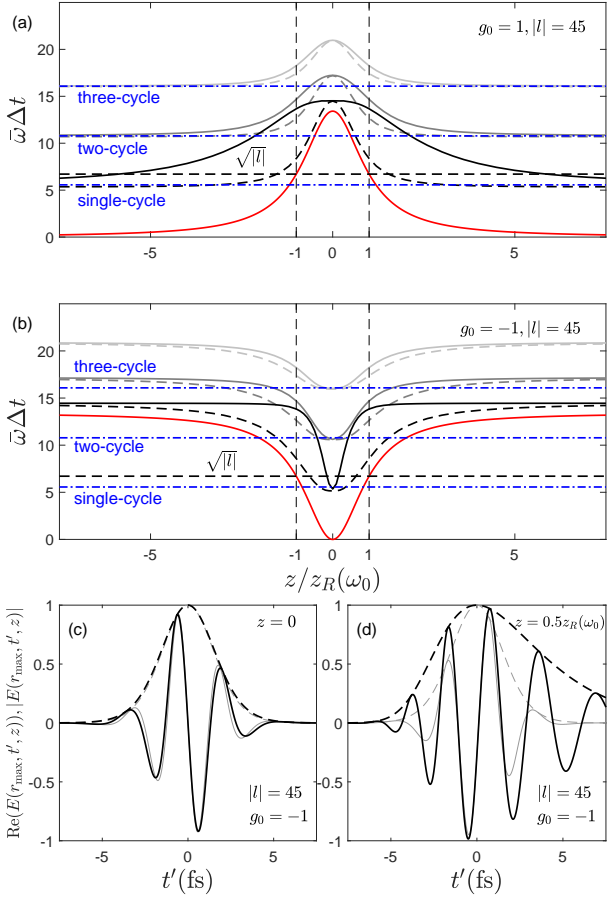}
\caption{\label{Fig5} (a) and (b) Black and gray solid curves: Change of the number of oscillations $\bar\omega\Delta t$ at the bright ring of UFVs during propagation evaluated numerically with Eqs. (\ref{F}), (\ref{E}), (\ref{DEBYE}), with $Z_R(\omega)=Z_R(\omega_0)(\omega/\omega_0)^{g_0}$ and $\hat A(\omega)= (\omega/\omega_0)^{\alpha-1/2}\exp(-\alpha\omega/\omega_0)$ with $\alpha=14.25$ (single-cycle pulse), $\alpha=57.11$ (two-cycle pulse) and $\alpha=128.30$ (three cycle pulse), and $|l|$=45. In (a) $g_0=1$ and in (b) $g_0=-1$. Solid red curve: Lower bound in inequality (\ref{LOWER3}). Dashed curves: Number of oscillations $\bar\omega\Delta t$ evaluated with transform-limited durations $\Delta t = 2/\Delta\omega$. Dash-dotted horizontal blue lines: $\bar\omega\Delta t$ for the source pulses. Dashed horizontal line: Isodiffracting lower bound $\sqrt{|l|}$. (c) and (d) Pulse shapes at the bright ring at the indicated distances (black curves) of the source single-cycle pulse (gray curves).}
\end{figure}

The change of the number of oscillations of the UFV upon propagation, as in the introductory example Figs. \ref{Fig2}(g-i), can be explained as a consequence of the existence of the $z$-varying lower bound in inequality (\ref{LOWER3}).

The solid black and gray curves in Fig. \ref{Fig5}(a) represent $\bar\omega\Delta t$ at the bright ring of the UFV as a function of propagation distance for the input UFVs $E_L(r,t)e^{il\varphi}= A(t)[\sqrt{2}r/S(\omega_0)]^{|l|}e^{-r^2/S(\omega_0)^2}e^{il\varphi}$ with $g_0=1$ because $Z_R(\omega)$ is such that $S(\omega)=S(\omega_0)$ is independent of frequency. The source spectra $\hat A(\omega)$ are chosen to represent Gaussian-like pulses $A(t)$ of one, two and three oscillations, whose values of $\omega_0\Delta t$ are represented as dashed-dotted blue lines for reference. As seen the number of oscillations of the respective UFVs (black, dark gray and light gray) remain at any propagation distance above the lower bound, represented as a red curve. The lower bound acts a kind of effective barrier that requires a significant broadening of the input single-cycle and two-cycle UFVs. A similar situation, but reversing focal region and far field, is given with $g_0=-1$, as illustrated in Fig. \ref{Fig5}(b). The input UFV is given by Eqs. (\ref{LENS1}) and (\ref{LENS2}) with $Z_R(\omega)$ such that $\Theta(\omega)=\Theta(\omega_0)$, and hence $s(\omega)=s(\omega_0)$ at focus, are independent of frequency, and with $\hat A(\omega)$ such that $A(t)$ comprises one, two and three oscillations, as above. Regardless of how low is $\omega_0\Delta t$ of the source pulse (dashed-dotted lines), the number of oscillations of the synthesized UFV on the focusing system and as it is directed towards the focus (black, dark gray and light gray curves) is above the lower bound (red curve), and it is only when the lower bound diminishes in the focal region that this particular UFV compresses to a duration also allowed by the lower bound in this region, and always above or equal to the duration of $A(t)$. For longer input durations, as the three-cycle UFV in both examples, these pulse shape changes during propagation are less and less pronounced.

\begin{figure}[!]
\centering \includegraphics*[width=8.5cm]{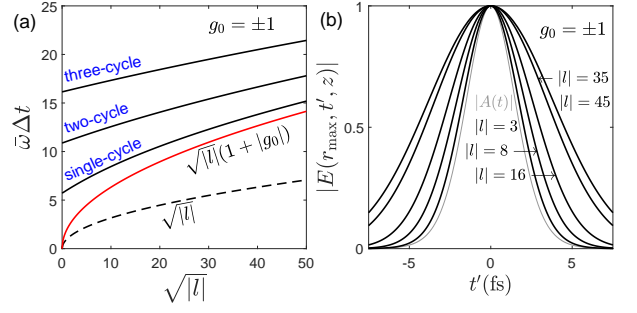}
\caption{\label{Fig6} (a) Number of oscillations $\bar\omega\Delta t$ at the bright ring at the focal plane for $g_0=1$ and at the far field for $g_0=-1$ with the same input conditions as in Fig. \ref{Fig5} as a function of the imprinted topological charge. The red curve is the lower bound in the inequality (\ref{LOWER3}) and the dashed black curve the lower bound for isodiffracting UFVs. (b) Broadened amplitudes for various values of $|l|$ at the focal plane for $g_0=1$ and at the far field ofr $g_0=-1$, compared to the source amplitude.}
\end{figure}

The proportionality of the lower bound to $\sqrt{|l|}$ entails an increase of the number of oscillations with $|l|$ at any particular location $z$, in the same way as for isodiffracting UFVs \cite{PORRAS1}. For the same input UFVs as in Figs. \ref{Fig5}(a) and (b), the number of oscillations and the lower bound $\sqrt{l|}(1+|g_0|)$ are represented in Fig. \ref{Fig6}(a) as functions of $|l|$ at the focal plane for $g_0=1$ and at the far field for $g_0=-1$. The existence of a lower bound monotonically increasing with $|l|$ at each particular location $z$ imposes the increase of the number of oscillations with increasing magnitude of the imprinted topological charge with respect to those of $A(t)$. As the frequency shifts are small, the envelopes shown in Fig. \ref{Fig6}(b) are increasingly broadened at the focus for $g_0=1$ and at the far field for $g_0=-1$ (they are identical) when increasing $|l|$ compared to the envelope of $A(t)$. Interestingly, this effect at an important location such as the focal plane is more pronounced than with isodiffracting UFVs, and should be observable in experiments even with values of $|l|$ below ten.

These above two examples also illustrate what we will refer to as ``locally compressed" states of OAM or UFVs, understood as UFVs whose number of oscillations is locally below the lower bound $\sqrt{|l|}$ for isodiffracting UFVs, represented in Fig. \ref{Fig5}(a) and (b) as dashed black lines. Indeed the bound $\sqrt{|l|}$ for isodiffracting UFVs continues to play a prominent role for general UFVs with $g_0\neq 0$ with axially varying pulse shape. Note that for $0<|g_0|\le 1$, the mean value of the minimum and maximum values of the lower bound, $\sqrt{|l|}(1-|g_0|)$ and $\sqrt{|l|}(1+|g_0|)$, respectively, along the propagation is just the isodiffracting value $\sqrt{|l|}$. Thus, as in Fig. \ref{Fig5}(a) for a source with $g_0=1$ and single-cycle $A(t)$, the UFV at the far field with $\bar\omega\Delta t < \sqrt{|l|}$ (black dashed line) can be regarded to be such a compressed UFV because it necessarily increases its number of oscillations to a value $\bar\omega\Delta t > \sqrt{|l|}$. Vice versa, for the source with $g_0=-1$ and single-cycle $A(t)$, it is possible to create, as in Fig. \ref{Fig5}(b), a UFV with $\bar\omega\Delta t <\sqrt{|l|}$ in the focal region, but it immediately broadens to $\bar\omega \Delta t > \sqrt{|l|}$. These locally compressed UFVs are located about the minimum of the lower bound in each case, but can only implemented in practice, as justified below, when the minimum lower bound $\sqrt{|l|}(1-|g_0|)$ is either in the far field (focusing system) or at the focus i. e., with sources with $|g_0|\le 1$, and can be optimally implemented with $g_0=\pm 1$ because the minimum lower bounds vanish. In these two cases, the value of $\bar\omega\Delta t$ at the far field or at the focus can reach its minimum practical value for the given source spectrum, as in Figs. \ref{Fig5}(a) and (b), where the pulse shape at the bright ring is almost identical to $A(t)$, except for the small red shift at the focus, as observed in Fig. \ref{Fig5}(c).

In principle, these compressed states OAM could survive, as allowed by the lower bound, the whole far field, $|z|>z_R(\omega_0)$, or the whole focal region, $|z|<z_R(\omega_0)$. However, as a general feature, they are much more localized axially, e. g., they exist only in a fraction of the far field in Fig. \ref{Fig5}(a) or in a fraction of the focal region in Fig. \ref{Fig5} (b), because of a new effect arising from the dependence of the Rayleigh distance with frequency, namely, the dispersions introduced by Gouy's phase and wave front mismatch. Unlike isodiffracting UFVs, a $\omega$-dependent Rayleigh distance introduces non-uniform spectral phases $-(|l|+1)\tan^{-1}[z/z_R(\omega)]$, strongly enhanced for large $|l|$, and $\omega r_{\rm max}^2/2cR(\omega,z)$, as the UFV approaches the focal region from outside, or immediately off-focus, that broadens and distorts the temporal pulse shape, as in Fig. \ref{Fig5}(d) at one half the Rayleigh distance, even if its bandwidth is similar to that of the compressed state at $z=0$. The limiting effect of these dispersions on the axial length of the compressed state is clear by comparing the pulse durations, depicted as the black curves, and the dashed curves in Figs. \ref{Fig5}(a) and (b). These dashed curves represent the number of oscillations that the pulse would have without Gouy's phase and wave front mismatch dispersions, evaluated as if the pulse would remain almost Gaussian-shaped with uniform spectral phases from the relation $\bar\omega\Delta t \simeq  \bar\omega 2/\Delta \omega$, where $\Delta\omega$ is the computed bandwidth. This comparison evidences that the length of the compressed states is reduced from almost the entire far field and almost the entire focal region to a small fractions of them by the effect of these dispersions in the respective cases of $g_0=1$ and $g_0=-1$.

\begin{figure}[!]
\centering \includegraphics*[width=8.5cm]{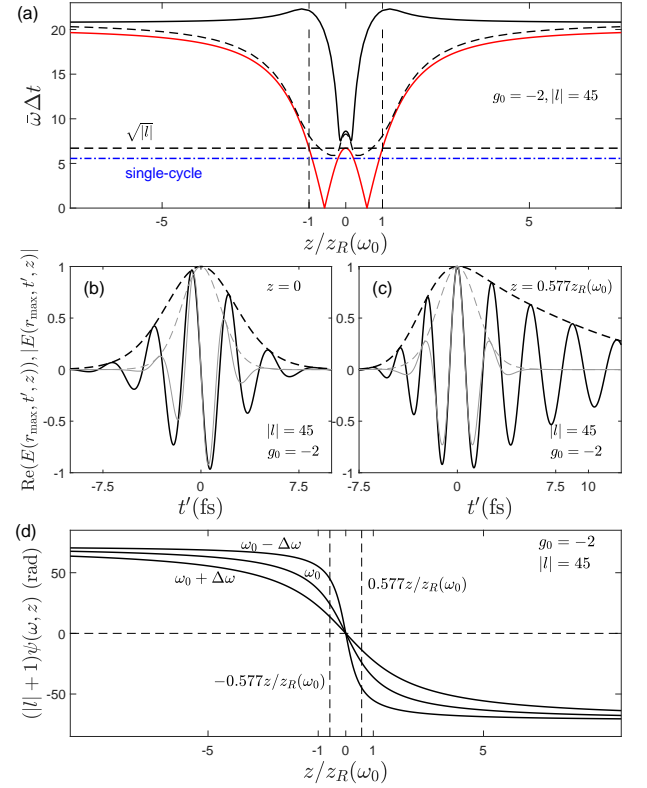}
\caption{\label{Fig7} (a) Black curve: Change of the number of oscillations $\bar\omega\Delta t$ at the bright ring of UFVs during propagation evaluated numerically with Eqs. (\ref{F}), (\ref{E}), (\ref{DEBYE}), with $Z_R(\omega)=Z_R(\omega_0)(\omega/\omega_0)^{g_0}$, $g_0=-2$, $\hat A(\omega)= (\omega/\omega_0)^{\alpha-1/2}\exp(-\alpha\omega/\omega_0)$,  $\alpha=14.25$ (single-cycle pulse) and $|l|$=45. Solid red curve: Lower bound in inequality (\ref{LOWER3}). Dashed gray curve: Number of oscillations $\bar\omega\Delta t$ evaluated with transform-limited durations $\Delta t = 2/\Delta\omega$. Dash-dotted horizontal blue line: $\bar\omega\Delta t$ for the source pulse. Dashed horizontal line: Isodiffracting lower bound $\sqrt{|l|}$. (b) and (c) Pulse shapes at the bright ring at the indicated distances (black curves) of the source single-cycle pulse (gray curves). (d) Gouy's phases of the monochromatic LG beams of the source carrier frequency $\omega_0$, and of frequencies $\omega_0+\Delta\omega$ and $\omega_0-\Delta\omega$ at the edge of the source spectrum.}
\end{figure}

With sources characterized by $|g_0|>1$, as that used in \cite{HOFF}, the situation is worse to the purpose of focusing to as short as possible pulse in a compressed UFV. First, the mean value of the maximum and minimum number of oscillations along the propagation is $\sqrt{|l|}(1+|g_0|)/2$, above the isodiffracting lower bound $\sqrt{|l|}$. Second, the points $z_b$ where the lower bound vanishes are not at infinity but somewhere outside the focal region for $g_0>1$, and not at the focus but somewhere in the focal region for $g_0<-1$, as illustrated in Fig. \ref{Fig7}. The possible compressed states located about $z_b$ are then broadened and distorted by Gouy's phase and wave front mismatch dispersions. With the same input conditions as in Fig. \ref{Fig5}(a) except that $g_0=-2$, the number of oscillations of the UFV (black curve) is seen in Fig. \ref{Fig7}(a) to be enormously increased on the focusing system and while it is directed towards the focus compared to the source single-cycle pulse (dashed-dotted blue line), as imposed by the lower bound (red curve). Removing artificially Gouy's phase and wave front mismatch dispersions, the UFV would focus about $z_b=0.577 z_R(\omega_0)$ into a compressed state (dashed gray curve) with number of oscillations below $\sqrt{|l|}$ (dashed horizontal line). The pulse at focus in Fig. \ref{Fig7}(b) is significantly broadened compared to $A(t)$, as imposed by the lower bound, the pulse shape at $z_b$ in Fig. \ref{Fig7}(c) is even more broadened and distorted as a result of dispersion, and at any propagation distance $\bar\omega\Delta t$ is above $\sqrt{|l|}$. Figure \ref{Fig7}(d) helps to visualize Gouy's phase dispersion, i. e., the different values of Gouy's phase for different spectral components, which is the main origin of the distortion for high topological charges.

To finish, it should be clear that reaching the minimum number of oscillations of the source pulse $A(t)$, i. e., the blue dash-dotted line in Fig. \ref{Fig7}(a), is not impossible because the lower bound actually vanishes at $z_b$, but only very difficult in practice. It would require measuring the non-uniform spectral phases of the UFV at $(r_{\rm max}, z_b)$ and pre-compensating for them prior to the focusing system, namely, and introducing spectral phases $(|l|+1)\tan^{-1}[z_b/z_R(\omega)]$ opposite to Gouy's phase and $-\omega r_{\rm max}^2/2cR(\omega,z_b)$ opposite to front mismatch for each particular frequency. This pre-compensation is however specific to the particular point $(r_{\rm max}, z_b)$ and the resulting UFV would be a locally compressed state of OAM about $z_b$.

\section{Conclusions}

In conclusion, we have conducted an analytical and numerical study of the free-space propagation features of general Laguerre-Gauss ultrafast vortices similar to those generated in experiments from femtosecond laser sources, whose Rayleigh distance is generally spectrally varying, as characterized by the $g_0$-factor. The first conclusion to keep in mind is that the simple view of an approximately invariable pulse envelope  modulated in space by a diffracting Laguerre-Gauss beam, which is widely taken for granted, but is only acceptable for few-cycle Gaussian beams with any reasonable $g_0$ factor \cite{PORRAS6,PORRAS10}, is far from describing the actual propagation characteristics of ultrafast vortices. Ultrafast vortices experience similar small frequency shifts at their bright ring as ultrafast Gaussian beams of the same value of $g_0$ at their center, but additionally they experience non-trivial changes in the pulse shape that depend on $|l|$ and $g_0$ during propagation, these changes being more pronounced as $|l|$ and $|g_0|$ are higher and as the ultrafast vortex is shorter.

Instead of studying in detail the characteristics of particular models of ultrafast vortices, we have extracted general laws underlying the behavior of all them and that explain the above phenomena. We have found an upper bound to the bandwidth relative to the carrier frequency, and a lower bound to the duration relative to the carrier period, i. e., to the number of oscillations, of the pulse at the most energetic ring, bounds that are satisfied by all synthesizable ultrafast vortices in experiments, and that generalize the bounds recently described for isodiffracting ultrafast vortices \cite{PORRAS1}. The lower bound to the number of oscillations, as given by the right hand side of inequality (\ref{LOWER3}) remains proportional to the lower bound $\sqrt{|l|}$ for isodiffracting ultrafast vortices, with axial modulations whose locations depends on $g_0$ and whose maximum and minimum average value is $\sqrt{|l|}$ for $|g_0|\le 1$, or the higher average value $\sqrt{|l|}(1+|g_0|)$ for $|g_0|>1$.

The existence of this lower bound explains the increase of the duration with increasing $|l|$ of the synthesized ultrafast vortex from the duration expected with the available bandwidth, this broadening being similar to that already described for isodiffracting ultrafast vortices \cite{PORRAS2}, and the axial modulation of the lower bound explains the changes in the duration of the vortex upon propagation.

The lower bound $\sqrt{|l|}$ for isodiffracting ultrafast vortices can be violated locally in what we have called locally compressed states of orbital angular momentum about the axial minima of the lower bound, but these states can be implemented in practice only with sources with $0<g_0\le 1$ at the far field, or with sources with $-1 \le g_0 < 0$ at the focal plane. The optimum condition to create a compressed state of orbital angular momentum in the focal plane is a source with $g_0=-1$ because the minimum lower bound vanishes at this plane. The term ``locally" stresses here that these states can only survive a small fraction of the Rayleigh distance because of Gouy's phase and wave front mismatch dispersions with $g_0\neq 0$ that strongly broaden and distort the pulse immediately off-focus.

Recent research has the revealed the dependence on the $g_0$-factor, characterizing variation with frequency of the beam parameters, of a variety of carrier-envelope-phase sensitive phenomena of interaction of few-cycle pulses with matter, such as electron acceleration with radially polarized pulses, high-harmonic and attosecond pulse generation, or electron photoemission from few-cycle pulses without orbital angular momentum \cite{HOFF,ZHANG2,JOLLY}. Given the nontrivial dependence of the propagation features of ultrafast vortices with the topological charge $l$ and the $g_0$-factor, this work stresses further the importance of measuring the $g_0$ factor of the femtosecond laser source in use for an adequate design and interpretation of experiments involving few-cycle pulses with orbital angular momentum, and for improvement and control of their applications.

\section*{acknowledgements}

This research was funded by Spanish Ministerio de Econom\'{\i}a y Competitividad, grant numbers PGC2018-093854-B-I00 and FIS2017-87360-P.

\end{document}